\documentclass[a4paper,11pt]{article}

\usepackage{amssymb}
\usepackage[english]{babel}

\newcommand{\qeda}{\hspace{10mm}\hfill $\square$}

\newtheorem{theorem}{Theorem}

\newtheorem{proposition}[theorem]{Proposition}

\begin{document}

\title{Local existence and uniqueness for a\\ semi-linear accretive wave equation}

\author{
\normalsize\textsc{H. Faour\footnote{Cermics, Paris-Est-ENPC,
ParisTech, 6 et 8 avenue Blaise Pascal, Cit\'e Descartes
Champs-sur-Marne, 77455 Marne-la-Vall\'ee Cedex 2, France. E-mail:
faourh@cermics.enpc.fr.\newline \indent ${}^1$ \hskip-.1cm A. Z. Fino
and M. Jazar, LaMA-Liban, AZM Research Center, Lebanese
University, P.O. Box 37, Tripoli Lebanon. E-mails:
afino@ul.edu.lb, mjazar@ul.edu.lb.\newline The first is supported
by the project ANR MICA (2006-2010).\newline The second author is
supported by a grant from the Lebanese National Research
Council.\newline The third authors is supported by a grant from
the Lebanese University.}, A. Z. Fino$^{1}$ and M. Jazar$^{1}$ }}
\vspace{20pt}

\maketitle

\centerline{\footnotesize{\bf{Abstract}}} \noindent{\small{We
study local existence and uniqueness in the phase space $H^\mu
\times H^{\mu-1}(\mathbb{R}^N)$ of the solution of the semilinear
wave equation $u_{tt}-\Delta u=u_t|u_t|^{p-1}$ for $p>1$.}}

\vskip 1cm


\section{Introduction and main results}\label{intro}
A very rich literature has been done on the semi-linear wave
equation
$$
u_{tt}-\Delta u=au_t\left| u_t\right| ^{p-1}+bu|u|^{q-1}
$$
with $a$, $b$, $p$ and $q$ are real numbers, $p,q\ge1$. When $a\le
0$ and $b=0$ then the damping term $au_t|u_t|^{p-1}$ ensures
global existence in time for arbitrary data (see, for instance,
Harraux and Zuazua \cite{HarauxZuazua} and Kopackova
\cite{Kopackova}). When $a\le 0$, $b>0$ and $p>q$ or when $a\le
0$, $b>0$ and $p=1$ then one can cite, for instance, Georgiev and
Todorova \cite{GeorgievTodorova} and Messaoudi \cite{Messaoudi},
that show the existence of global solutions under negative energy
condition.

The first to consider the case $a>0$ was Haraux \cite{Haraux1992}
(with $b=0$ on bounded domain), who construct blowing up solutions
for arbitrary small initial data. See also Jazar and Kiwan
\cite{JazarKiwan1} and the references therein for the same
equation on bounded domain.

In this paper we consider the case $a=1$ and $b=0$, i.e. the
semi-linear accretive wave equation
\begin{equation}\label{11wave}
\left\{
\begin{array}{ll}
\displaystyle{u_{tt}-\Delta
u=u_t|u_t|^{p-1}}&\displaystyle{x\in {\mathbb{R}^N},t>0,}\\
\\
\displaystyle{u(0,x)=u_0(x),\;u_t(0,x)=u_1(x)}&\displaystyle{x\in
{\mathbb{R}^N}.}
\end{array}
\right. \end{equation} To our knowledge, no local existence result
was done for this equation. The phase space to consider here is
$Y^\mu:=H^\mu\times H^{\mu-1}(\mathbb{R}^N)$, and we are looking
to find conditions on the nonlinearity $p$ and the order $\mu$ so
that we have local existence. Due to our method, based on the use
of Strichartz estimates (Proposition \ref{GV}) and bounds on a
power of a function in a Sobolev space $\|h^p\|_{H^s}$ by the norm
of the initial function $\|h\|_{H^r}$, we need that $p$ or $\mu$
to be integer. This is done in the following two theorems

\begin{theorem}\label{local1}(Case $p$ integer)\\
Let $p\in \mathbb{N}\backslash\{0,1\}$ and
$\mu\in\mathbb{R}\cap[1,\infty)$ such that
\begin{equation}\label{mu}
\left\{
\begin{array}{ll}
\displaystyle{1<p<\infty}&\displaystyle{\mbox{if }\mu\geq
1+\frac{N}{2},}\\
&\\
\displaystyle{1<p<\frac{N+4-2\mu}{N+2-2\mu}}&\displaystyle{\mbox{if
}1\leq\mu<1+\frac{N}{2}.}
 \end{array}
 \right.
\end{equation}
For $(u_0,v_0)\in Y^\mu$, there exists a maximal time $T_{max}>0$
and a unique solution $(u,u_t)\in C^0([0,T_{max});Y^\mu)$ of
problem (\ref{11wave}). Moreover, if $T_{\max}<\infty,$ we have
$\|u(t)\|_{H^\mu}+\|u_t(t)\|_{H^{\mu-1}}\rightarrow\infty$ as
$t\rightarrow T_{\max}.$

\end{theorem}

One can compare to the case $a=0$ and $b=1$, the restriction on
$p$ is the same by taking $\mu=2$, see for instance
\cite{MerleZaag2003, ShatahStruwe}).

In the previous theorem, $p$ is integer. In the following theorem
this is no longer the case. However, the dimension must be less
than 3 or equal. This is due to Proposition \ref{lemma1} in which,
we obtain $L^\infty$-estimates on the wave kernel. In what follows
denote by $Y^{2,\infty}:=W^{2,\infty}\times
W^{1,\infty}(\mathbb{R}^N)$.

\begin{theorem}\label{local2}(Case $p$ real)\\
Let $1\le N\le 3$, $\mu\in(1,2)\cap \mathbb{N}^*$ and $p\in
(1,\infty)\cap [\mu-1,\infty)$.  Then for all $(u_0,v_0)\in
Y^\mu\cap Y^{2,\infty}$ there exists $T>0$ and a unique solution
$(u,u_t)\in C^0([0,T];(L^\infty\cap H^\mu)\times L^\infty\cap
H^{\mu-1}(\mathbb{R}^N))$ of (\ref{11wave}).
\end{theorem}

\section{Preliminary notations and results}
In this section we use the notations used by
\cite{GinibreVeloJFA1995}. For $x\in X$ a normed vector space we
denote by $\|x;X\|$ the norm of $x$, and for $(x,y)\in X\times Y$
then, naturally, $\|(x,y);X\times Y\|=\|x;X\|+\|y;Y\|$. Finally,
for $q\in [1,+\infty)$ define the norm
$\|f;L^q(0,T;X)\|^q:=\int_0^T\|f(t);X\|^qdt$ with the usual one
for $q=+\infty$.

Consider the inhomogeneous wave equation in
$\mathbb{R}\times\mathbb{R}^N$
\begin{equation}\label{lwave}
\left\{\begin{array}{l} u_{tt}-\Delta u=f\\
u(0,x)=u_0(x),\,\,u_t(0,x)=v_0(x).
\end{array}\right.
\end{equation}
We define the operator $\sigma:=(-\Delta)^{1/2}$, which could be
defined as $\sigma u
(x)=\mathcal{F}^{-1}(|\xi|\mathcal{F}(u)(\xi))(x)$, and
$K(t):=\sigma^{-1}\sin\sigma t$, $\dot{K}(t):=\cos\sigma t$. The
solution of (\ref{lwave}) could be written as $u=\theta+\omega$,
where $\theta$ is the solution of the homogeneous equation with
the same initial data
\begin{equation}\label{lihwave}
\left\{\begin{array}{l} \theta_{tt}-\Delta \theta=0\\
\theta(0,x)=u_0(x),\,\,\theta_t(0,x)=v_0(x)
\end{array}\right.
\end{equation}
namely
$$\theta(t)=\dot{K}(t)u_0+K(t)v_0\mbox{ and }\theta_t(t)=\Delta K(t)u_0+\dot{K}(t)v_0$$
that we denote by $H(t)U_0=(\theta(t), \theta_t(t))$ where
$U_0:=(u_0,v_0)$.
 And $\omega$
is the solution of the inhomogeneous equation with zero initial data
\begin{equation}\label{lhwave}
\left\{\begin{array}{l} \omega_{tt}-\Delta \omega=f\\
\omega(0,x)=\omega_t(0,x)=0.
\end{array}\right.
\end{equation}
The solution of (\ref{lhwave}) could be written, for $t\ge 0$, as
$$\omega(t)=\int_0^tK(t-s)f(s)ds=K\star f(t) \mbox{ and }
\omega_t(t)=\int_0^t\dot{K}(t-s)f(s)ds=\dot{K}\star f(t).$$ The
initial data $U_0$ will be taken in the phase space $Y^\mu$ for
$\mu\in\mathbb{R}$ where $H^\mu$ is the homogeneous Sobolev space
(See \cite{Triebel}). We will use the following ``simplified
version" of the generalized Strichartz inequality
\cite{StrichartzDuke77}:

\begin{proposition}\label{GV}(Proposition 3.1 of
\cite{GinibreVeloJFA1995})
Let $\rho_1,\rho_2,\mu\in\mathbb{R}$ and $q_1,q_2\ge 2$ and let
the following condition be satisfied
\begin{equation}
\label{erhoq}\rho_1-1/q_1=\mu=1-(\rho_2-1/q_2).
\end{equation}
Then
\begin{enumerate}
\item $\|H(\cdot)U_0;L^{q_1}(\mathbb{R}, Y^{\rho_1})\|\le
C\|U_0;Y^\mu\|$.

\item For any interval $I=[0,T)$, $T\le\infty$, then
$$\|(\omega(\cdot),\omega_t(\cdot));L^{q_1}(I, Y^{\rho_1})\|\le
C\|f;L^{\overline{q}_2}(I,H^{-\rho_2})\|.$$
\end{enumerate}
The constants $C$ are independent of the interval $I$.
\end{proposition}

For the proof of Theorem \ref{local2} we need the following proposition
\begin{proposition}\label{lemma1}(\cite[Lemma
3.4]{HosonoOgawa} and \cite{Nishihara})\\ Let  $1\leq N\leq3,$
$f\in W^{1,\infty}(\mathbb{R}^N)$ and $g\in
L^\infty(\mathbb{R}^N).$ Then, we have
$$\|\dot{K}(t)f;L^\infty(\mathbb{R}^N)\|\leq
\max(1,t)\|f;W^{1,\infty}(\mathbb{R}^N)\|,\quad t>0,$$
$$\|K(t)g;L^\infty(\mathbb{R}^N)\|\leq t\|g;L^\infty(\mathbb{R}^N)\|,\quad t>0.$$
\end{proposition}

We will need also to deal with the Sobolev norm of a power of a
function. The following is a direct consequence of
\cite{Bachelot}:

\begin{proposition}\label{product}
Let $p\in \mathbb{N}\backslash\{0,1\}$, $s>-N/2$, $s\ne N/2$ and
$\nu(s,p):=\sup\{0,(N/2-s)(p-1)/p\}$. Then, for a nonnegative
function $f\in H^{s+\nu(s,p)}(\mathbb{R}^N)$, $f^p\in
H^s(\mathbb{R}^N)$ and there exists a positive constant $C$ such
that
$$\|f^p;H^s(\mathbb{R}^N)\|\le C\|f;H^{s+\nu(s,p)}(\mathbb{R}^N)\|^p.$$
\end{proposition}

Next, we give the Gagliardo-Nirenberg inequality which is a direct consequence
of \cite[Theorem 1.3.4]{CazHar} (see also \cite[Theorem 9.3]{Friedman}).
\begin{proposition}\label{G-N}(Gagliardo-Nirenberg)\\
Let $q,r$ be such that $1\leq q,r\leq\infty$, and let $j,m$ be integers,
$0\leq j<m$. Let $a\in[j/m,1]$ $(a<1\;\mbox{if}\; m-j-N/r\;\mbox{is an integer}
\geq0)$, and let $p$ be given by
$$\frac{1}{p}=\frac{j}{N}+a\left(\frac{1}{r}-\frac{m}{N}\right)+(1-a)\frac{1}{q}.$$
For $f\in L^q(\mathbb{R}^N)$ such that $ D^\alpha f\in L^r(\mathbb{R}^N)$
with $|\alpha|=m$, we have $ D^\alpha f\in L^p(\mathbb{R}^N)$ with $|\alpha|=j$ and
there exists a positive constant $C$ such that
$$\sum_{|\alpha|=j}\|D^\alpha f\|_{L^p}\leq C\left(\sum_{|\alpha|=m}\|D^\alpha f\|_{L^r}\right)^a\|f\|^{1-a}_{L^q}.$$
\end{proposition}

For the second theorem we need

\begin{proposition}\label{power2}
Assume that $\mu\in (1,2)\cup \mathbb{N}^*$ and $p\in
\mathbb{R}\cap (1,\infty)\cap [\mu-1,\infty)$. For a nonnegative
function $f\in L^{\infty}(\mathbb{R}^N)\cap
H^{\mu-1}(\mathbb{R}^N)$, $f^p\in H^{\mu-1}(\mathbb{R}^N)$ and
there exists a positive constant $C$ such that
$$\|f^p\|_{H^{\mu-1}(\mathbb{R}^N)}\leq C\|f\|^{p-1}_{L^\infty(\mathbb{R}^N)}
\|f\|_{H^{\mu-1}(\mathbb{R}^N)}.$$
\end{proposition}
\noindent\textbf{Proof.} \underline{First case: $\mu\in(1,2)$}.
Using the mean value theorem and the following equivalent norm of
$||\cdot||_{H^{\mu-1}}$ (see \cite[Theorem 7.48 page 214]{Adams}),
we have:
\begin{eqnarray*}
  \|f^p\|^2_{H^{\mu-1}}&:=&\|f^p\|^2_{L^2}+\int_{\mathbb{R}^N}\int_{\mathbb{R}^N}
  \frac{\left(f^p(x)-f^p(y)\right)^2}{|x-y|^{N+2\mu-2}}\,dx\,dy \\
   &=&\|f^p\|^2_{L^2}+\int_{\mathbb{R}^N}\int_{\mathbb{R}^N}\frac{\left(pz^{p-1}
   (f(x)-f(y))\right)^2}{|x-y|^{N+2\mu-2}}\,dx\,dy   \\
   &\leq&p^2\|f^{p-1}\|^2_{L^\infty}\left[\|f\|^2_{L^2}+\int_{\mathbb{R}^N}
   \int_{\mathbb{R}^N}\frac{\left(f(x)-f(y)\right)^2}{|x-y|^{N+2\mu-2}}\,dx\,dy\right]\\
   &\leq&p^2\|f\|_{L^\infty}^{2(p-1)}\|f\|^2_{H^{\mu-1}},
\end{eqnarray*}
where $\min(f(x),f(y))<z<\max(f(x),f(y))$ for every
$x,y\in\mathbb{R}^N.$\\
\underline{Second case: $\mu\in\mathbb{N}^*$}. We have
\begin{equation} \label{def1}
\|f^p\|_{H^{\mu-1}}:=\left(\sum_{0\leq|\alpha|\leq
\mu-1}\|D^\alpha f^p\|_{L^2}^2\right)^{1/2}.
\end{equation}
As
$$D^{\alpha}(f^p)=\sum_{|\beta_1|+\cdots +|\beta_{\alpha}|=
|\alpha|}
C_{|\alpha|,|\beta_1|,\cdots,|\beta_\alpha|}f^{p-|\alpha|}
D^{\beta_1} f\dots D^{\beta_\alpha} f,$$ we conclude that
\begin{eqnarray*}
\|D^\alpha f^p\|_{L^2} &\leq& C\|f\|_{L^\infty}^{p-|\alpha|}
\sum_{|\beta_1|+\cdots +|\beta_{\alpha}|=|\alpha|} \|D^{\beta_1}
f\dots D^{\beta_\alpha} f\|_{L^2}\\
&\leq& C\|f\|_{L^\infty}^{p-|\alpha|}  \sum_{|\beta_1|+\cdots
+|\beta_{\alpha}|=|\alpha|} \|D^{\beta_1}
f\|_{L^{\frac{2|\alpha|}{|\beta_1|}}}\cdots \| D^{\beta_\alpha}
f\|_{L^{\frac{2|\alpha|}{|\beta_\alpha|}}}
\end{eqnarray*}
thanks to H\"older's inequality. Using Proposition \ref{G-N}, we obtain
\begin{eqnarray*}
\|D^\alpha f^p\|_{L^2}& \leq &C\|f\|_{L^\infty}^{p-|\alpha|}
\sum_{|\beta_1|+\dots +|\beta_{\alpha}|=|\alpha|} \left[
\sum_{|\beta|=|\alpha|}\|D^{\beta}
f\|_{L^2}\right]^{|\beta_1|/|\alpha|}
\|f\|_{L^\infty}^{1-|\beta_1|/|\alpha|}\cdots\\
&& \hskip 2cm\left[ \sum_{|\beta|=|\alpha|}\|D^{\beta}
f\|_{L^2}\right]^{|\beta_\alpha|/|\alpha|}
\|f\|_{L^\infty}^{1-|\beta_\alpha|/|\alpha|}\\
&\leq& C\|f\|_{L^\infty}^{p-1} \|D^\alpha f\|_{L^2},
\end{eqnarray*}
which gives, using $(\ref{def1})$, the desired
estimates.$\hfill\square$

\section{Proof of theorems \ref{local1} and \ref{local2}}

\noindent\textbf{Proof of Theorem \ref{local1}}. Without loss of
generality, we may assume that the initial data $(u_0,v_0)$ is of
compact support. By finite speed of propagation, the solution
$(u,u_t)$ is also of compact support. This allows us to use
Proposition \ref{GV} on bounded intervals $[0,T]$ with usual
Sobolev spaces instead of homogeneous Sobolev spaces (since for
compactly supported distributions the norms are equivalent, see
\cite{Triebel}).

Now, we write (\ref{11wave}) as
\begin{equation}\label{12wave}
\left\{\begin{array}{l} U_t-A U=F(U)\\
\\
U(0,x)=U_0(x)\in Y^\mu,
\end{array}\right.
\end{equation}
where $U=(u,v)$, $A=\left[\begin{array}{ll}0&\hskip.25cm I\\
\Delta&\hskip.25cm 0\end{array}\right]$ and $F(U):=\left[\begin{array}{c}0\\
v|v|^{p-1}\end{array}\right]$. With theses notations, local
existence for (\ref{11wave}) is equivalent to local existence for
(\ref{12wave}), and this is equivalent to the following integral
equation
\begin{equation}\label{13wave}
\left\{\begin{array}{l} \mbox{Find }T>0\mbox{ and a unique
solution},\\ \\ U\in C^0([0,T];Y^\mu) \mbox{ of }
U(t)=H(t)U_0+L(U)(t),
\end{array}\right.
\end{equation}
where $H(t)U_0=\left[\begin{array}{c}\dot{K}(t)u_0+K(t)v_0\\
\Delta K(t)u_0+\dot{K}(t)v_0\end{array}\right]$ and
$L(U)(t)=\left[\begin{array}{c} [K\star v^p](t)\\
{}[\dot{K} \star v^p](t)\end{array}\right]$, where $v^p$ denotes
$v|v|^{p-1}$.

In order to use Fixed Point Theorem, let us introduce the
following metric space
$$X:=\{\phi=(\phi_1,\phi_2)\in L^q(0,T;Y^\rho)\mbox{ s.t. }\|\phi-HU_0;
L^q(0,T;Y^\rho)\|\le \lambda\}$$ where $T$ and $\lambda$ are
positive constants and $\rho$ and $q$ satisfy (\ref{erhoq}). These
constants will be fixed later.

For $\phi,\psi\in X$ denote by
$$\|\phi;X\|:=\|\phi-HU_0;L^q(0,T;Y^\rho)\|$$
($HU_0$ denotes the function $t\mapsto H(t)U_0$) and the natural
induced distance
$$d(\phi,\psi):=\|\phi-\psi;L^q(0,T;Y^\rho)\|.$$
Finally define the map $\Phi$ on $X$ by
$\Phi(U)(t):=H(t)U_0+L(U)(t)$.

\noindent\underline{First step: $X$ is invariant under $\Phi$}.
Let $U\in X$, by proposition \ref{GV} we have
$$\|\Phi(U);X\|\le \|L(U);L^q(0,T;Y^\rho)\|\le
C\|v^p;L^{\overline{q_2}}(0,T;H^{-\rho_2})\|$$
for all $(q_2,\rho_2)$ satisfying (\ref{erhoq}). Take
$q_2:=+\infty$, hence $\rho_2:=1-\mu$. Then
$$\|\Phi(U);X\|\le C\|v^p;L^1(0,T;H^{\mu-1})\|.$$
By proposition \ref{product}
$$\|v^p;H^{\mu-1}\|\le C\|v;H^{\mu-1+\nu(\mu-1,p)}\|^p,$$
thus
\begin{equation}\label{inv1}\|\Phi(U);X\|\le
C\|v;L^p(0,T;H^{\mu-1+\nu(\mu-1,p)})\|^p.\end{equation}

If $\mu\ge 1+N/2$ then $\nu(\mu-1,p)=0$ and choosing
$\varepsilon=1$, $q=\infty$ and $\rho=\mu$ then (\ref{inv1}) gives
\begin{eqnarray}\|\Phi(U);X\|&\le&
C\|v;L^p(0,T;H^{\mu-1})\|^p=C\int_0^T\|v;H^{\mu-1}\|^p\nonumber \\
&\le&\label{inv2} CT\|v;L^\infty(0,T;H^{\mu-1})\|^p \le
CT^\varepsilon\|U;L^q(0,T;Y^\rho)\|^p .\end{eqnarray}

If $\mu<1+N/2$, define
$\varepsilon:=1-p\nu(\mu-1,p)=1-(p-1)(1+N/2-\mu)\in(0,1)$ by
(\ref{mu}). Then $\mu+\nu(\mu-1,p)-\frac{1-\varepsilon}p=\mu$.
Using H\"older's inequality we have
$$\|v;L^p(0,T;H^{\mu-1+\nu(\mu-1,p)})\|^p\le
CT^\varepsilon\|v;L^{\frac
p{1-\varepsilon}}(0,T;H^{\mu-1+\nu(\mu-1,p)})\|^p.$$ By choosing
$\rho=\mu+\nu(\mu-1,p)$ and $q=p/(1-\varepsilon)$, and using
proposition \ref{GV}, inequality (\ref{inv1}) gives
$$
\|\Phi(U);X\|\le CT^\varepsilon \|v;L^q(0,T;H^{\rho-1})\|\le
CT^\varepsilon \|U;L^q(0,T;Y^\rho)\|^p.
$$
We see that, in both cases one has
\begin{equation}\label{inv3}
\|\Phi(U);X\|\le CT^\varepsilon \|U;L^q(0,T;Y^\rho)\|^p,
\end{equation}
with $\rho$ and $q$ satisfying (\ref{erhoq}). Thus we have
$$\|\Phi(U);X\|\le CT^\varepsilon \|U;L^q(0,T;Y^\rho)\|^p\le
CT^\varepsilon \left[\|U;X\|+\|HU_0;L^q(0,T;Y^\rho)\|  \right]^p,
$$
and, as $\rho$ and $q$ satisfy (\ref{erhoq}), by proposition
\ref{GV} we have
$$\|\Phi(U);X\|\le CT^\varepsilon \left[\|U;X\|+\|U_0;Y^\mu\|
\right]^p.
$$
Therefore, $X$ is invariant by $\Phi$ if $T$ and $\lambda$ are
such that
\begin{equation}\label{CXT1}
CT^\varepsilon(\lambda+\|U_0;Y^\mu\|)^p\le \lambda.
\end{equation}

\noindent\underline{Second step: $\Phi$ is a contraction on
$X$}.\\
This is mainly the same ideas. Let $U,V\in X$. Then
$$d(\Phi(U),\Phi(V))=\|L(U)-L(V);L^q(0,T;Y^\rho)\|\le
C\|U_2^p-V_2^p;L^{\overline{q_2}}(0,T;H^{-\rho_2})\|$$ for
$\rho_2$ and $q_2$ satisfying (\ref{erhoq}). Take $q_2:=+\infty$,
we get $\rho_2:=1-\mu$ and then
$$d(\Phi(U),\Phi(V))\le C\|U_2^p-V_2^p;L^1(0,T;H^{\mu-1})\|.$$

Now, we write $U_2^p-V_2^p=(U_2-V_2)P(U_2,V_2)$ where $P$ is a
homogeneous polynomial of degree $p-1$. Using proposition
\ref{product} and the convexity of the exponential function to
obtain
\begin{eqnarray*}d(\Phi(U),\Phi(V))&\le&
C\|U_2-V_2;L^p(0,T;H^{\mu-1+\nu(\mu-1,p)})\|\\
&&\hskip-1cm\times\left[\|U_2;L^p(0,T;H^{\mu-1+\nu(\mu-1,p)})\|^{p-1}
+\|V_2;L^p(0,T;H^{\mu-1+\nu(\mu-1,p)})\|^{p-1}\right].
\end{eqnarray*}
By the same analysis, and a similar calculation as in the first
step, we get
$$d(\Phi(U),\Phi(V))\le Cd(U,V)T^\varepsilon [\lambda+\|U_0;Y^\mu\|]^{p-1}.$$
Finally, $\Phi$ is a contraction on $X$ if $\lambda$ and $T$
satisfy
\begin{equation}\label{CXT2}
CT^\varepsilon[\lambda+\|U_0;Y^\mu\|]^{p-1}<1.
\end{equation}
By choosing $\lambda$ and $T$ satisfying (\ref{CXT1}) and
(\ref{CXT2}), Fixed Point Theorem ensures existence and
uniqueness.

\noindent\underline{Third step: continuity of the solution}.\\
We have obtained existence and uniqueness of a solution
$U=(u,u_t)\in X\subset L^q(0,T;Y^\rho)$ where $\rho$ and $q$
satisfy (\ref{erhoq}). Let's show that $U\in
L^{q'}(0,T;Y^{\rho'})$ for any $\rho'$ and $q'$ satisfying
(\ref{erhoq}). This is a similar calculation to the first step.
Indeed,
\begin{eqnarray*}\|U;L^{q'}(0,T;Y^{\rho'})\|&\le&
C\|v^p;L^1(0,T;Y^{\mu-1})\|+\|U_0;Y^\mu\|\\
 &\le&
CT^\varepsilon\|U;L^{q}(0,T;Y^{\rho})\|^p
+\|U_0;Y^\mu\|<\infty.\end{eqnarray*}

In particular $U\in L^\infty(0,T,Y^\mu)$. Now, using
(\ref{13wave}), for $0\le t_0\le t$, we have
\begin{eqnarray*}
\|U(t)-U(t_0);L^\infty(t_0,t;Y^\mu)\le \|U(t)-H(t-t_0)U_0\|
+\|U(t_0)-H(t-t_0)U_0\|.
\end{eqnarray*}
As in the first step
\begin{eqnarray*}
\|U(t)-H(t-t_0)U_0;L^\infty(t_0,t;Y^\mu)\|&\le&
C(t-t_0)^\varepsilon\|U(t);L^q(t_0,t;Y^\rho)\|^p,
\end{eqnarray*}
and the strong continuity of the $C_0$-group $H$ associated to the
wave equation implies that
$$\lim_{t\to t_0}\|H(t-t_0)U_0-U_0;L^\infty(t_0,t;Y^\mu)\|=0.$$
Therefore, $U\in C^0([0,T];Y^\mu)$. \qeda

\bigskip

\noindent{\bf Proof of Theorem \ref{local2}.} Using Proposition
\ref{lemma1}, the proof is similar to that of Theorem \ref{local1}
where we use Proposition \ref{power2} instead of Proposition
\ref{product}. $\hfill\square$

\bibliographystyle{plain}
\bibliography{c:/bib/mybib}

\begin{thebibliography}{10}

\bibitem{Adams}
Robert~A. Adams.
\newblock {\em Sobolev spaces}.
\newblock Academic Press [A subsidiary of Harcourt Brace Jovanovich,
  Publishers], New York-London, 1975.
\newblock Pure and Applied Mathematics, Vol. 65.

\bibitem{Bachelot}
A.~Bachelot.
\newblock Formes quadratiques compatibles dans les espaces de type $l^p$.
\newblock {\em Compte-rendu CNRS}, (8406), 1984.

\bibitem{CazHar}
Thierry Cazenave and Alain Haraux.
\newblock {\em Introduction aux probl\`emes d'\'evolution semi-lin\'eaires},
  volume~1 of {\em Math\'ematiques \& Applications (Paris) [Mathematics and
  Applications]}.
\newblock Ellipses, Paris, 1990.

\bibitem{Friedman}
Avner Friedman.
\newblock {\em Partial differential equations of parabolic type}.
\newblock Prentice-Hall Inc., Englewood Cliffs, N.J., 1964.

\bibitem{GeorgievTodorova}
Vladimir Georgiev and Grozdena Todorova.
\newblock Existence of a solution of the wave equation with nonlinear damping
  and source terms.
\newblock {\em J. Differential Equations}, 109(2):295--308, 1994.

\bibitem{GinibreVeloJFA1995}
J.~Ginibre and G.~Velo.
\newblock Generalized {S}trichartz inequalities for the wave equation.
\newblock {\em J. Funct. Anal.}, 133(1):50--68, 1995.

\bibitem{Haraux1992}
A.~Haraux.
\newblock Remarks on the wave equation with a nonlinear term with respect to
  the velocity.
\newblock {\em Portugal. Math.}, 49(4):447--454, 1992.

\bibitem{HarauxZuazua}
A.~Haraux and E.~Zuazua.
\newblock Decay estimates for some semilinear damped hyperbolic problems.
\newblock {\em Arch. Rational Mech. Anal.}, 100(2):191--206, 1988.

\bibitem{HosonoOgawa}
Takafumi Hosono and Takayoshi Ogawa.
\newblock Large time behavior and {$L\sp p$}-{$L\sp q$} estimate of solutions
  of 2-dimensional nonlinear damped wave equations.
\newblock {\em J. Differential Equations}, 203(1):82--118, 2004.

\bibitem{Kopackova}
M.~Kop{\'a}{\v{c}}kov{\'a}.
\newblock Remarks on bounded solutions of a semilinear dissipative hyperbolic
  equation.
\newblock {\em Comment. Math. Univ. Carolin.}, 30(4):713--719, 1989.

\bibitem{MerleZaag2003}
Frank Merle and Hatem Zaag.
\newblock Determination of the blow-up rate for the semilinear wave equation.
\newblock {\em Amer. J. Math.}, 125(5):1147--1164, 2003.

\bibitem{Messaoudi}
Salim~A. Messaoudi.
\newblock Blow up and global existence in a nonlinear viscoelastic wave
  equation.
\newblock {\em Math. Nachr.}, 260:58--66, 2003.

\bibitem{Nishihara}
Kenji Nishihara.
\newblock {$L\sp p$}-{$L\sp q$} estimates of solutions to the damped wave
  equation in 3-dimensional space and their application.
\newblock {\em Math. Z.}, 244(3):631--649, 2003.

\bibitem{ShatahStruwe}
Jalal Shatah and Michael Struwe.
\newblock Geometric wave equations.
\newblock 2:viii+153, 1998.

\bibitem{StrichartzDuke77}
Robert~S. Strichartz.
\newblock Restrictions of {F}ourier transforms to quadratic surfaces and decay
  of solutions of wave equations.
\newblock {\em Duke Math. J.}, 44(3):705--714, 1977.

\bibitem{Triebel}
Hans Triebel.
\newblock {\em Theory of function spaces}, volume~78 of {\em Monographs in
  Mathematics}.
\newblock Birkh\"auser Verlag, Basel, 1983.

\end{thebibliography}

\end{document}